\documentclass[12pt]{article}
\usepackage{amssymb}

\normalbaselines
\input tcilatex
\begin{document}

\author{Yuri A. Rylov}
\title{Incompatibility of the Copenhagen interpretation with quantum
mechanics formalism}
\date{Institute for Problems in Mechanics, Russian Academy of Sciences,\\
101-1, Vernadskii Ave., Moscow, 119526, Russia.\\
e-mail: rylov@ipmnet.ru\\
}
\maketitle

\begin{abstract}
It is proved the mathematical theorem, that the wave function describes the
statistical ensemble of particles, but not a single particle. Supposition,
that the wave function describes a single particle appears to be
incompatible with formalism of quantum mechanics.
\end{abstract}

Interest to interpretation of quantum mechanics was very large from the very
beginning of the quantum mechanics creation up to now. Some scientific
journals (Physics Today (1999), Uspechi Fizicheskich Nauk (2002)) organized
discussions devoted to problems of quantum measurements and their
interpretation. There are scientists \cite{B76,B70,B98}, which believe that
the wave function describes a statistical ensemble. There are scientists 
\cite{H55,P64,P97,N04}, which believe that the wave function describes a
single particle. There are scientists \cite{L64,L67,L74}, whose position is
intermediate. There is a lot of papers devoted to interpretation of quantum
mechanics. All discussions were produced on the verbal level. None of
researchers had not set the problem mathematically: Which of interpretations
does follow from the quantum mechanics formalism? or in negative form: Which
of interpretations is incompatible with the quantum mechanics formalism?
Such a statement of the problem seems to be very reasonable. However, the
question in such a form was not set. After mathematical solution of this
problem any discussion on the verbal level seems to be useless.

In this paper we prove a very important theorem, which claims that the wave
function may not describe an individual quantum particle. It describes
always a statistical ensemble of quantum particles. We shall show, that the
action $\mathcal{A}_{\mathrm{S}}$ for the Schr\"{o}dinger particle $\mathcal{%
S}_{\mathrm{S}}$ (the dynamic system described by the Schr\"{o}dinger
equation) turns into the action $\mathcal{A}_{\mathcal{E}\left[ \mathcal{S}_{%
\mathrm{cl}}\right] }$ for the statistical ensemble $\mathcal{E}\left[ 
\mathcal{S}_{\mathrm{cl}}\right] $ of free classical particles $\mathcal{S}_{%
\mathrm{cl}}$, when the quantum constant $\hbar \rightarrow 0$. Such a
transition is possible only in the case, \textit{when the wave function }$%
\psi $\textit{\ describes a statistical ensemble of quantum particles, but
not a single particle}.

For the free Schr\"{o}dinger particle $\mathcal{S}_{\mathrm{S}}$ the action
has the form 
\begin{equation}
\mathcal{S}_{\mathrm{S}}:\qquad \mathcal{A}_{\mathrm{S}}\left[ \psi ,\psi
^{\ast }\right] =\int \left\{ \frac{i\hbar }{2}\left( \psi ^{\ast }\partial
_{0}\psi -\partial _{0}\psi ^{\ast }\cdot \psi \right) -\frac{\hbar ^{2}}{2m}%
\mathbf{\nabla }\psi ^{\ast }\mathbf{\nabla }\psi \right\} dtd\mathbf{x}
\label{a1.2}
\end{equation}%
where $\psi =\psi \left( t,\mathbf{x}\right) $ is a complex one-component
wave function, $\psi ^{\ast }=\psi ^{\ast }\left( t,\mathbf{x}\right) $ is
the complex conjugate to $\psi $, and $m$ is the particle mass. It is
supposed that in the classical limit $\hbar \rightarrow 0$ the description
of the dynamic system $\mathcal{S}_{\mathrm{S}}$ becomes to be a classical
description of a free particle $\mathcal{S}_{\mathrm{cl}}$.

However, there are two different classical descriptions of the free
classical particle $\mathcal{S}_{\mathrm{cl}}$. The individual classical
particle $\mathcal{S}_{\mathrm{cl}}$ is described by the action%
\begin{equation}
\mathcal{A}_{\mathcal{S}_{\mathrm{cl}}}\left[ \mathbf{x}\right] =\int \frac{m%
}{2}\left( \frac{d\mathbf{x}}{dt}\right) ^{2}dt  \label{b1.1}
\end{equation}%
where $\mathbf{x}=\left\{ x^{1}\left( t\right) ,x^{2}\left( t\right)
,x^{3}\left( t\right) \right\} $.

Statistical ensemble $\mathcal{E}\left[ \mathcal{S}_{\mathrm{cl}}\right] $
of free classical particles $\mathcal{S}_{\mathrm{cl}}$ is described by the
action 
\begin{equation}
\mathcal{A}_{\mathcal{E}\left[ \mathcal{S}_{\mathrm{cl}}\right] }\left[ 
\mathbf{x}\right] =\int \frac{m}{2}\left( \frac{d\mathbf{x}}{dt}\right)
^{2}dtd\mathbf{\xi }  \label{a1.22}
\end{equation}%
where $\mathbf{x}=\left\{ x^{1}\left( t,\mathbf{\xi }\right) ,x^{2}\left( t,%
\mathbf{\xi }\right) ,x^{3}\left( t,\mathbf{\xi }\right) \right\} $.
Parameters $\mathbf{\xi =}\left\{ \xi _{1},\xi _{2},\xi _{3}\right\} $ label
elements (particles) of the statistical ensemble $\mathcal{E}\left[ \mathcal{%
S}_{\mathrm{cl}}\right] $. Both dynamic systems $\mathcal{S}_{\mathrm{cl}}$
and $\mathcal{E}\left[ \mathcal{S}_{\mathrm{cl}}\right] $ are classical.
However, $\mathcal{S}_{\mathrm{cl}}$ has six degrees of freedom (the order
of the system of the first order ordinary differential equations), whereas
the statistical ensemble $\mathcal{E}\left[ \mathcal{S}_{\mathrm{cl}}\right] 
$ has infinite number of the freedom degrees, because it consists of the
infinite number of the particles $\mathcal{S}_{\mathrm{cl}}$. The dynamic
system $\mathcal{E}\left[ \mathcal{S}_{\mathrm{cl}}\right] $ may be
interpreted as an ideal fluid without pressure. This fluid may be described
in terms of a wave function \cite{R99}. In this case the action (\ref{a1.22}%
) has the form%
\begin{equation}
\mathcal{A}\left[ \psi ,\psi ^{\ast }\right] =\int \left\{ \frac{ib}{2}%
\left( \psi ^{\ast }\partial _{0}\psi -\partial _{0}\psi ^{\ast }\psi
\right) -\frac{b^{2}}{2}\mathbf{\nabla }\psi ^{\ast }\mathbf{\nabla }\psi +%
\frac{b^{2}}{8\rho }\left( \rho ^{2}\mathbf{\nabla }s_{\alpha }\mathbf{%
\nabla }s_{\alpha }+\left( \mathbf{\nabla }\rho \right) ^{2}\right) \right\}
d^{4}x  \label{a1.23}
\end{equation}%
where $b$ is a real constant $b\neq 0$, $\psi =\left( _{\psi _{2}}^{\psi
_{1}}\right) $ is a two-component complex wave function and $\psi ^{\ast
}=\left( \psi _{1}^{\ast },\psi _{2}^{\ast }\right) $ is the complex
conjugate to $\psi $.%
\begin{equation}
\rho =\psi ^{\ast }\psi ,\qquad s_{\alpha }=\frac{\psi ^{\ast }\sigma
_{\alpha }\psi }{\rho },\qquad \alpha =1,2,3  \label{a1.24}
\end{equation}%
and $\sigma _{\alpha }$, $\alpha =1,2,3$ are the Pauli matrices. In the case
when the flow is irrotational, the wave function $\psi $ may be chosen
one-component. In this case $s_{\alpha }=$const and the action (\ref{a1.23})
turns into the action 
\begin{equation}
\mathcal{A}\left[ \psi ,\psi ^{\ast }\right] =\int \left\{ \frac{ib}{2}%
\left( \psi ^{\ast }\partial _{0}\psi -\partial _{0}\psi ^{\ast }\psi
\right) -\frac{b^{2}}{2}\mathbf{\nabla }\psi ^{\ast }\mathbf{\nabla }\psi +%
\frac{b^{2}}{8\rho }\left( \mathbf{\nabla }\rho \right) ^{2}\right\} d^{4}x
\label{a1.25}
\end{equation}

Let us investigate, into what classical dynamic system ($\mathcal{S}_{%
\mathrm{cl}}$ or $\mathcal{E}\left[ \mathcal{S}_{\mathrm{cl}}\right] $)
turns the dynamic system $\mathcal{S}_{\mathrm{S}}$ in the limit $\hbar
\rightarrow 0$?

The quantum constant $\hbar $ is a parameter of the dynamic system (\ref%
{a1.2}). As a rule, a change of a parameter of a dynamic system does not
change the number and the character of dynamic equations. The number of the
freedom degrees does not changes also. The dynamic system $\mathcal{S}_{%
\mathrm{S}}$ has infinite number of the freedom degrees, and we should
expect that at $\hbar \rightarrow 0$ the dynamic system $\mathcal{S}_{%
\mathrm{S}}$ turns into $\mathcal{E}\left[ \mathcal{S}_{\mathrm{cl}}\right] $%
, which also has infinite number of the freedom degrees, but not into $%
\mathcal{S}_{\mathrm{cl}}$, which has six degrees of freedom.

However, at $\hbar =0$ the description by means of the action (\ref{a1.2})
degenerates, and one should consider the limit $\hbar \rightarrow 0$ of the
description by means of the action (\ref{a1.2}). To obtain this limit, we
make a change of variables 
\begin{equation}
\psi \rightarrow \Psi _{b}=\left\vert \psi \right\vert \exp \left( \frac{%
\hbar }{b}\log \frac{\psi }{\left\vert \psi \right\vert }\right) ,\qquad
\psi =\left\vert \Psi _{b}\right\vert \exp \left( \frac{b}{\hbar }\log \frac{%
\Psi _{b}}{\left\vert \Psi _{b}\right\vert }\right)  \label{a1.9}
\end{equation}%
where $b\neq 0$ is some real constant. After this change of variables the
action (\ref{a1.2}) turns into 
\begin{equation}
\mathcal{A}_{\mathcal{S}_{\mathrm{q}}}\left[ \Psi _{b},\Psi _{b}^{\ast }%
\right] =\int \left\{ \frac{ib}{2}\left( \Psi _{b}^{\ast }\partial _{0}\Psi
_{b}-\partial _{0}\Psi _{b}^{\ast }\cdot \Psi _{b}\right) -\frac{b^{2}}{2m}%
\mathbf{\nabla }\Psi _{b}^{\ast }\mathbf{\nabla }\Psi _{b}-\frac{\hbar
^{2}-b^{2}}{2m}\left( \mathbf{\nabla }\left\vert \Psi _{b}\right\vert
\right) ^{2}\right\} dtd\mathbf{x}  \label{a1.10}
\end{equation}%
The transformation (\ref{a1.9}) is analytical for any values of parameters $%
b $ and $\hbar $, except for the case, when $\hslash =0$ or $b=0$. The
constant $b$ is arbitrary, and it always can be chosen $b\neq 0$. The value $%
\hbar =0$ is not considered, because in this case the action (\ref{a1.2}),
as well as the transformation (\ref{a1.9}) degenerate. For all values of $%
\hbar \neq 0$ the dynamic systems (\ref{a1.2}) and (\ref{a1.10}) are
equivalent. At $\hbar \rightarrow 0$ the dynamic system (\ref{a1.10}) does
not degenerate, it turns into the dynamic system $\mathcal{E}^{\prime }\left[
\mathcal{S}_{\mathrm{cl}}\right] $ 
\begin{equation}
\mathcal{A}_{\mathcal{E}^{\prime }\left[ \mathcal{S}_{\mathrm{cl}}\right] }%
\left[ \Psi _{b},\Psi _{b}^{\ast }\right] =\int \left\{ \frac{ib}{2}\left(
\Psi _{b}^{\ast }\partial _{0}\Psi _{b}-\partial _{0}\Psi _{b}^{\ast }\cdot
\Psi _{b}\right) -\frac{b^{2}}{2m}\mathbf{\nabla }\Psi _{b}^{\ast }\mathbf{%
\nabla }\Psi _{b}+\frac{b^{2}}{2m}\left( \mathbf{\nabla }\left\vert \Psi
_{b}\right\vert \right) ^{2}\right\} dtd\mathbf{x}  \label{a1.12}
\end{equation}%
which may be considered as the limit of the action (\ref{a1.2}) at $\hbar
\rightarrow 0$. The action (\ref{a1.12}) is a partial case of the action (%
\ref{a1.22}), because (\ref{a1.12}) coincides with (\ref{a1.25}), if one
takes into account that $\left\vert \Psi _{b}\right\vert =\left\vert \psi
\right\vert =\sqrt{\rho }$. Thus, dynamic system $\mathcal{E}^{\prime }\left[
\mathcal{S}_{\mathrm{cl}}\right] $ is a special case of the dynamic system $%
\mathcal{E}\left[ \mathcal{S}_{\mathrm{cl}}\right] $

But independently of, whether or not dynamic systems $\mathcal{E}^{\prime }%
\left[ \mathcal{S}_{\mathrm{cl}}\right] $ and $\mathcal{E}\left[ \mathcal{S}%
_{\mathrm{cl}}\right] $ coincide, the dynamic system (\ref{a1.12}) cannot
coincide with the dynamic system (\ref{b1.1}), because the dynamic system (%
\ref{b1.1}) has six degrees of freedom, whereas the dynamic system (\ref%
{a1.12}) has infinite number of the freedom degrees. It means that the 
\textit{wave function may not describe a single particle}, and the
Copenhagen interpretation and other QM interpretations, founded on the
statement, that the wave function describes a single particle, may not be
used. In particular, such phenomena as superluminal interaction in the EPR
experiment and many-worlds interpretation \cite{E57,DG73} appear to be
impossible as founded on the statement, that the wave function describes an
individual particle.

\end{document}